\documentclass[twocolumn,prd]{revtex4}% Physical Review D
\usepackage{graphicx}
\begin{document}
\title{Quasinormal modes prefer supersymmetry ?}
\author{
Yi Ling$^{1,2}$\footnote{Email: yling@itp.ac.cn}, Hongbao
Zhang$^{3,4}$\footnote{Email: hongbaozhang@pku.edu.cn}}
\affiliation{%
${}^1$ CCAST (World Laboratory), P.O. Box 8730, Beijing
   100080, China}
\affiliation{%
${}^2$ Institute of Theoretical Physics,
 Chinese Academy of Sciences,
 P.O.Box 2735, Beijing 100080, China}
\affiliation{%
${}^3$ School of Physics, Peking University, Beijing, 100871,
China}
\affiliation{%
${}^4$ Institute of Applied Mathematics, Academy of Sciences,
Beijing, 100080, China}

\begin{abstract}
One ambiguity in loop quantum gravity is the appearance of a free
parameter which is called Immirzi parameter. Recently Dreyer has
argued that this parameter may be fixed by considering the
quasinormal mode spectrum of black holes, while at the price of
changing the gauge group to $SO(3)$ rather than the original one
$SU(2)$. Physically such a replacement is not quite natural or
desirable. %In this rapid communication we argue that the Immirzi
%parameter may be fixed very well if we consider the simplest
%supersymmetric extension of loop quantum gravity. Our result
%strongly indicates that $N=1$ supersymmetry might be a favorable
%nature of spacetime.
In this paper we study the relationship between the black hole
entropy and the quasi normal mode spectrum in the loop
quantization of $N=1$ supergravity. We find that a %there  a
single value of the Immirzi parameter agrees with the
semiclassical expectations as well. %in both cases.
But in this case the lowest supersymmetric representation
dominates, fitting well with the result based on statistical
consideration. This suggests that, so long as fermions are
included in the theory, supersymemtry may be favored for the
consistency of the low energy limit of loop quantum gravity.

\end{abstract}\maketitle
\section{Introduction}

Loop quantum gravity (LQG) has in the past years been further
developed and made great success in the study of the quantum
theory of geometry (please see \cite{Rev} for recent review). One
remarkable result is that with the power of spin networks the area
and volume spectra can be derived and characterized by discrete
values\cite{SN2}. Nevertheless, in loop quantum gravity there is
an important parameter $\gamma$ called Immirzi parameter
unfixed\cite{Immirzi}, implying that given a two dimensional
surface intersected by an edge with label $J$ of a spin network,
the area of this surface can be determined only up to a freely
adjustable parameter,
\begin{equation}
A(J)=8\pi\gamma l_p^2\sqrt{J(J+1)}\label{as},
\end{equation}
where $l_p$ is the Plank length (taking $\sqrt{\hbar}$ in
geometrical units) and $J$ takes positive half-integer.
Consequently it is also due to the Immirzi parameter that the
famous factor $1\over 4$ in Bekenstein-Hawking entropy
formula,\begin{equation} S_{BH}={A\over
4l^2_p},\label{BH}\end{equation} could not be uniquely determined
when applying quantum geometry to derive the statistical entropy
of black holes\cite{EBH}. This has been viewed as the main
unsatisfactory point of this approach for some years.

Recently, motivated by Hod's work on the asymptotic quasinormal
modes of black holes\cite{Hod}, Dreyer has proposed a novel way to
fix the Immirzi parameter in loop quantum gravity\cite{Dreyer}.
The basic idea and the main results can be summed as follows.
Having studied the asymptotic behavior of the quasinormal modes of
a Schwarzschild black hole with the mass $M$, Hod conjectured that
the real part of the highly damped quasinormal mode frequencies
$\omega_{QNM}$ asymptotically approaches to a fixed quantity
\cite{Hod}
\begin{equation}
\omega_{QNM}=\frac{\ln 3}{8\pi M}.
\end{equation} This conjecture has recently been confirmed
analytically by Motl\cite{Motl}. Hod also proposed that, if we
assume the Bohr's correspondence principle is applicable to black
holes, such an asymptotic frequency of the quasinormal modes
 should be consistent with the one of the quanta induced by the minimum
change $\Delta A$ in the quantized area of the event horizon
\cite{Hod}. Going further, Dreyer made a striking observation that
the Immirzi parameter may be fixed by applying this idea to loop
quantum gravity. As a consequence, it is shown in \cite{Dreyer}
that
\begin{equation}
\gamma=\frac{\ln 3}{2\sqrt{2}\pi}.
\end{equation}
However, at the same time it also forces \begin{equation}
J_{min}=1,
\end{equation}
where $J_{min}$ denotes the lowest
possible spin for the representation of the gauge group. This is not the answer as we %naively
expect based on the statistical principle. As in the case of
quantum general relativity, the gauge group is $SU(2)$ and its
representation is labelled by positive half-integer, we expect the
minimum value $J_{min}$ should be ${1\over 2}$, rather than one.
The above observation by Dreyer seemingly indicates that we have
to either sacrifice $SU(2)$ as the gauge group and replace it with
$SO(3)$, or find a mechanism to explain why the expected most
important contribution from $J=1/2$ edges is suppressed. The
former option is not quite physically natural or desirable. Thus
some attempts to save $SU(2)$ as the relevant gauge group in loop
quantum gravity have been made (for instance see
\cite{Corichi,Swain}), but there additional physical
considerations or assumptions are needed.

In this paper %rapid communication
we intend to present another quite simple but elegant scheme to
fix the Immirzi parameter by considering the supersymmetric
extension of loop quantum gravity, %In this case it turns out the
%Immirzi parameter can also be fixed well,
but more importantly, we find in this case $J_{min}$ is forced to
take the minimum value of the representation of gauge group
$Osp(1|2)$ which is perfectly consistent with the most probable
value as we expect based on the statistical consideration.

\section{Area spectrum in $N=1$ supergravity and black hole entropy}
$N=1$ supergravity in Ashtekar-Sen variables was originally given
by Jacobson in \cite{Jacobson} and its canonical quantization has
been extensively studied in\cite{SUGRA,LS}. In this section we
only recall the crucial results that we need here; for more
details on this subject please see\cite{SUGRA,LS}. Similar to the
case in quantum general relativity, a generalized notion of
supersymmetric spin networks can be introduced to construct the
Hilbert space of canonical supergravity, while a key different
ingredient from the ordinary one is that the edges in spin
networks now are labelled by the representations of the supergroup
$Osp(1|2)$ rather than $SU(2)$. Such a replacement gives rise to a
different expression for the area spectrum at quantum mechanical
level. More explicitly, if a surface is intersected by an edge of
the network carrying the label $J$, then its area is given by
\cite{LS}
\begin{equation}
A(J)=8\pi\tilde{\gamma} l_p^2\sqrt{J(J+\frac{1}{2})},\label{sas}
\end{equation}
where $J$ still takes half-integer but
 labels $Osp(1|2)$ representation and $\tilde{\gamma}$ denotes the Immirzi
 parameter in supergravity. If we simply set $\gamma=\tilde{\gamma}$ and compare it with the area
 spectrum (\ref{as}) in ordinary loop quantum gravity,
 we find that the discrepancy is so tiny that it can %always
be ignored for large $J$, while for small values of $J$  these two
sorts of spectra in principle
 should be distinguishable and perhaps yield quite different physics. We will soon see it does so for the statistical
 entropy of black holes as discussed below.

 The derivation of statistical entropy of black holes in loop quantum gravity
 has been thoroughly studied and we refer to \cite{EBH,Smolin} for details. Here we simply
 argue that the strategy can be directly extended to consider the entropy of black holes (at least for
 Schwarzschild black holes) in
 the context of supergravity.  The only thing that should be stressed is that the dimension of
 the Hilbert space associated to a puncture with spin $J$ on the event horizon now is \cite{LS}
 \begin{equation}
 D(J)=4J+1.
 \end{equation}
Therefore, the area of its event horizon with $N$ punctures may be
obtained as
 \begin{equation}
A=N8\pi\tilde{\gamma}
l_p^2\sqrt{J_{min}(J_{min}+\frac{1}{2})},\label{area}
 \end{equation}
 and the number of microscopic quantum states contributes the black hole an entropy
 with
 \begin{equation}
S=N\ln(4J_{min}+1).\label{entropy}
 \end{equation}
where we have used the fact that given a black hole with fixed
horizon, the most probable distribution of spins associated to the
punctures is the configuration in which all the punctures are
labelled by the minimum value of spin, denoted as $J_{min}$.
Combining Eq.(\ref{area}) and Eq.(\ref{entropy}), we find the
black hole entropy is proportional to the area of the event
horizon, i.e.
\begin{equation}
S=\frac{\ln(4J_{min}+1)}{8\pi\tilde{\gamma}
l_p^2\sqrt{J_{min}(J_{min}+\frac{1}{2})}}A.\label{EA}
\end{equation}
Furthermore, comparing this with Bekenstein-Hawking formula
(\ref{BH}) yields the following relation between the Immirzi
parameter and the minimum value of spin,
\begin{equation}\frac{\ln(4J_{min}+1)}{8\pi\tilde{\gamma}
\sqrt{J_{min}(J_{min}+\frac{1}{2})}}={1\over
4}.\label{eq1}\end{equation}

Next we follow the strategy advocated in \cite{Dreyer} to fix the
Immirzi parameter by considering the asymptotic quasinormal modes
of black holes, but present a more reasonable value of $J_{min}$.
\section{Fixing the Immirzi parameter by quasinormal mode spectrum}
It is argued in \cite{Hod,Dreyer} that according to Bohr's
correspondence principle, the increase of the black hole mass
should be ascribed to a quanta with the energy $\hbar
\omega_{QNM}$ absorbed by the black hole, i.e.
\begin{equation}
\Delta M=\hbar \omega_{QNM}.
\end{equation}
On the other hand, since
\begin{equation}
A=16\pi M^2,
\end{equation}
such an increase in the mass of the black hole induces a
corresponding increase in the area of the event horizon
\begin{equation}
\Delta A=32\pi M\Delta M=4\ln 3l_p^2.\label{deltaA}
\end{equation}
At quantum mechanical level, this kind of increase in the area of
the horizon corresponds to the appearance of another edge labelled
with $J_{min}$ in the supersymmetric spin network, thus
\begin{equation}
\Delta A=A(J_{min}).
\end{equation}
Plugging it into Eq.(\ref{deltaA}) we have
\begin{equation}
\tilde{\gamma}=\frac{\ln
3}{2\pi\sqrt{J_{min}(J_{min}+\frac{1}{2})}}.\label{Immirzi}
\end{equation}
Therefore, Eq.(\ref{Immirzi}) together with Eq.(\ref{eq1})
uniquely fixes the value of the Immirzi parameter to be
\begin{equation}
\tilde{\gamma}=\frac{\ln 3}{\sqrt{2}\pi}.
\end{equation}
At the same time it also determines \begin{equation}
J_{min}=\frac{1}{2}.
\end{equation}
This is a remarkable result because it is consistent with our
argument based on statistical principle, stating that the most
probable distribution should be the set of punctures on the
horizon uniformly labelled by the fundamental representation of
supergroup $Osp(1|2)$ which takes the minimum positive value of
$J$. Here we do not need any other extra assumption or condition.
It only seems implying that the supersymmetry might be a favorable
nature of spacetime.
\section{Discussion}
We conclude this paper by pointing out a very interesting relation
between two area spectra respectively obtained in quantum general
relativity and the simplest supergravity, namely Eq.(\ref{as}) and
Eq.(\ref{sas}). Note in quantum gravity, we do not have any reason
to argue that those two Immirzi parameters should be related in
any manner. Therefore these two spectra in principle have nothing
to do with each other. One naive but natural choice is just
setting two Immirzi parameters equal as we assumed in section two
such that at each level labelled by the same $J$ the spectrum has
only a small shift after the supersymmetry is concerned. In
particular, such a shift is vanishing as $J$ approaches to
infinity. However, this picture has greatly changed when the
semi-classical limit of these two theories is concerned. The
asymptotic quasinormal modes of black holes do provide us a
strategy to fix the Immirzi parameter so that these two sorts of
spectra are comparable. Quite strikingly, here we found our
previous intuition is not quite correct. As a matter of fact, two
Immirzi parameters are not equal but
\begin{equation}
\tilde{\gamma}=2\gamma.\end{equation} Therefore, the area spectrum
in canonical supergravity can be rewritten as
\begin{equation}
A^{sg}(J)=8\pi\gamma l_p^2\sqrt{2J(2J+1)}\equiv 8\pi\gamma
l_p^2\sqrt{J_{eff}(J_{eff}+1)}.\label{sas2}
\end{equation}
Now it's evident that the area spectrum is the same expression as
the one in quantum general relativity except that $J_{eff}$ here
can only
take integer! That is to say, all the area eigenvalues with half-integers
in loop quantum gravity now are forbidden by the supersymmetry in $N=1$ supergravity.
%\footnote{Note also that, in the canonical
%supergravity, the dimension of the relevant Hilbert space takes
%$2J_{eff}+1$ similar to the one in quantum general relativity.}!

Corichi and Swain have suggested to suppress the contribution from
$J=1/2$ edges in loop quantum gravity by considering the coupling
of fermions or Pauli exclusion principle respectively. From this
point of view, we may also argue that we have presented a very
elegant scheme to do so, but the reason is simply due to the
supersymmetry. Interestingly enough, all these suggestions share
one common point, namely the involvement of fermions. As a result,
it's quite desirable to investigate their relations and show more
insight into this subject.

So far, we have only focused on Schwarzschild-like black holes. We
expect the similar consideration will be applicable to more
general black holes. At the same time, investigating the
supersymmetric extension of loop quantum gravity with $N>1$ is
also under progress.

\section*{Acknowledgement}
It's our pleasure to thank Olaf Dreyer, Yongge Ma and Lee Smolin
for correspondence and helpful discussions. Y. Ling is partly
supported by NSFC (No.10205002) and K.C.Wong Education Foundation,
Hong Kong.


\begin{thebibliography}{0}
%\cite{Thiemann:2001yy}
\bibitem{Rev}
L. Smolin,
%``How far are we from the quantum theory of gravity?,''
arXiv:hep-th/0303185; T. Thiemann,
%``Introduction to modern canonical quantum general relativity,''
arXiv:gr-qc/0110034;
%%CITATION = GR-QC 0110034;%%
 A. Ashtekar,
%``Quantum geometry and gravity: Recent advances,''
arXiv:gr-qc/0112038.
%%CITATION = GR-QC 0112038;%%

%%CITATION = HEP-TH 0303185;%%
%\cite{Rovelli:1994ge}
\bibitem{SN2}
C.~Rovelli and L.~Smolin,
%``Discreteness of area and volume in quantum gravity,''
Nucl.\ Phys.\ B {\bf 442}, 593 (1995) [Erratum-ibid.\ B {\bf 456},
753 (1995)] [arXiv:gr-qc/9411005].
%%CITATION = GR-QC 9411005;%%

%\cite{Immirzi:1996dr}
\bibitem{Immirzi}
G.~Immirzi,
%``Quantum gravity and Regge calculus,''
Nucl.\ Phys.\ Proc.\ Suppl.\  {\bf 57}, 65 (1997)
[arXiv:gr-qc/9701052].
%%CITATION = GR-QC 9701052;%%



%\cite{Rovelli:1996dv}
\bibitem{EBH}K. Krasnov, Phys.\ Rev.\ D. {\bf 55},3505(1997);
C.~Rovelli,
%``Black hole entropy from loop quantum gravity,''
Phys.\ Rev.\ Lett.\  {\bf 77}, 3288 (1996) [arXiv:gr-qc/9603063];
%%CITATION = GR-QC 9603063;%%
A.~Ashtekar, J.~Baez, A.~Corichi and K.~Krasnov,
%``Quantum geometry and black hole entropy,''
Phys.\ Rev.\ Lett.\  {\bf 80}, 904 (1998) [arXiv:gr-qc/9710007].
%%CITATION = GR-QC 9710007;%%

\bibitem{Hod} S. Hod, Phys.\ Rev.\ Lett. {\bf 81}, 4293(1998).
\bibitem{Dreyer} O. Dreyer, Phys.\ Rev.\ Lett.\ {\bf 90}, 081301(2003).
\bibitem{Motl} L. Motl, Adv.\ Theor.\ Math.\ Phys.\ {\bf 6},
1135 (2003).

\bibitem{Corichi} A. Corichi, Phys.\ Rev.\ D {\bf 67}, 087502(2003).
\bibitem{Swain} J. Swain, gr-qc/0305073.


%\cite{Jacobson:1987cj}
\bibitem{Jacobson}
T.~Jacobson,
%``New Variables For Canonical Supergravity,''
Class.\ Quant.\ Grav.\  {\bf 5}, 923 (1988).
%%CITATION = CQGRD,5,923;%%

%\cite{Gambini:1995db}
\bibitem{SUGRA}
R.~Gambini, O.~Obregon and J.~Pullin,
%``Towards a loop representation for quantum canonical supergravity,''
Nucl.\ Phys.\ B {\bf 460}, 615 (1996) [arXiv:hep-th/9508036];
%%CITATION = HEP-TH 9508036;%%
H.~Kunitomo and T.~Sano,
%``The Ashtekar's formulation for canonical $N = 2$ supergravity,''
Int.\ J.\ Mod.\ Phys.\ D {\bf 1}, 559 (1993) [Prog.\ Theor.\
Phys.\ Suppl.\  {\bf 114}, 31 (1993)];
%%CITATION = IMPAE,D1,559;%%
T.~Sano and J.~Shiraishi,
%``The Nonperturbative canonical quantization of the N=1 supergravity,''
Nucl.\ Phys.\ B {\bf 410}, 423 (1993) [arXiv:hep-th/9211104];
%%CITATION = HEP-TH 9211104;%%
T.~Sano,
%``The Ashtekar formalism and WKB wave functions of N=1, N=2 supergravities,''
arXiv:hep-th/9211103;
%%CITATION = HEP-TH 9211103;%%
K.~Ezawa,
%``Ashtekar's formulation for $N=1,2$ supergravities as "constrained" BF theories,''
Prog.\ Theor.\ Phys.\  {\bf 95}, 863 (1996)
[arXiv:hep-th/9511047].
%%CITATION = HEP-TH 9511047;%%

\bibitem{LS} Y. Ling and L. Smolin, Phys.\ Rev.\ D {\bf 61},
044008(2000).
%\cite{Smolin:1995vq}
\bibitem{Smolin}
L.~Smolin,
%``Linking topological quantum field theory and nonperturbative quantum gravity,''
J.\ Math.\ Phys.\  {\bf 36}, 6417 (1995) [arXiv:gr-qc/9505028].
%%CITATION = GR-QC 9505028;%%
\end{thebibliography}
\end{document}